\documentclass[12pt,aps,prd,tightenlines,groupedaddress,preprint,floatfix,nofootinbib]{revtex4-1}

\usepackage{graphicx}
\usepackage{amssymb}
\usepackage{epstopdf}

\newcommand{\PRE}[1]{{#1}} 

\def\be{\begin{equation}}
\def\ee{\end{equation}}
\def\bea{\begin{eqnarray}}
\def\eea{\end{eqnarray}}

\def\tev{\, {\rm TeV}}
\def\gev{\, {\rm GeV}}
\def\mev{\, {\rm MeV}}
\def\kev{\, {\rm keV}}

\newcommand{\gsim}{\lower.7ex\hbox{$\;\stackrel{\textstyle>}{\sim}\;$}}
\newcommand{\lsim}{\lower.7ex\hbox{$\;\stackrel{\textstyle<}{\sim}\;$}}

\begin{document}

\setlength{\baselineskip}{0.25in}

\setcounter{footnote}{0}
\setcounter{page}{1}
\setcounter{figure}{0}
\setcounter{table}{0}


\preprint{UH-511-1233-2014}

\title{\PRE{\vspace*{0.8in}}
Complementary Constraints on Light Dark Matter from Heavy Quarkonium Decays
\PRE{\vspace*{0.3in}}
}
\author{Nicolas Fernandez}
\affiliation{Department of Physics and Astronomy, University of Hawaii, Honolulu, HI 96822 USA
\PRE{\vspace*{.5in}}
}

\author{Jason Kumar}
\affiliation{Department of Physics and Astronomy, University of Hawaii, Honolulu, HI 96822 USA
\PRE{\vspace*{.5in}}
}

\author{Ilsoo Seong}
\affiliation{Department of Physics and Astronomy, University of Hawaii, Honolulu, HI 96822 USA
\PRE{\vspace*{.5in}}
}

\author{Patrick Stengel}
\affiliation{Department of Physics and Astronomy, University of Hawaii, Honolulu, HI 96822 USA
\PRE{\vspace*{.5in}}
}

\begin{abstract}
\PRE{\vspace*{.3in}}
We investigate constraints on the properties of light dark matter which can be obtained from analysis of invisible quarkonium decays at high intensity
electron-positron colliders in the framework of a low energy effective field theory. A matrix element analysis of all contact operators pertinent
for these meson decays allows for a model-independent calculation of associated dark matter-nucleon scattering and dark matter
annihilation cross sections. Assuming dark matter
couples universally to all quark flavors, we then obtain bounds on nucleon scattering which complement direct
dark matter detection searches. In contrast
to similar analyses of monojet searches at high energy colliders, B and charm factories are more suitable
probes of light dark matter interactions with
less massive mediators. Relevant bounds on dark matter annihilation arising from gamma ray searches of dwarf
spheroidal galaxies are also presented.

\end{abstract}


\maketitle

\section{Introduction}
Despite strong observational evidence for non-baryonic dark matter (DM) which interacts gravitationally~\cite{Kolb:1990}, the
detection of dark matter interactions with the Standard Model (SM) remains elusive. Many extensions of the SM predict dark matter candidates
which should leave signatures in the direct and indirect dark matter detection experiments, and at hadron colliders.
If the particles mediating dark matter-Standard Model interactions are much heavier than the energy scales involved, then the
constraints on dark matter interactions arising from these disparate detection strategies can be related
to each other in a model independent fashion via a generalized effective field theory (EFT) framework,
in which the details of the ultraviolet (UV) physics have been integrated out of the
Lagrangian~\cite{Goodman:2011,Hooper:2009,Cao:2011,Kumar:2013iva,Rajaraman:2013,Dreiner:2013,Buckley:2013,Rajaraman:2013-2,
Busoni:2014,Buchmueller:2014,Crivellin:2014,Busoni:2014-2,Alves:2014,Fedderke:2014},
 and dark matter-Standard Model interactions occur through contact operators.

Weakly interacting massive particles (WIMPs) are stable dark matter candidates predicted by many models of
physics beyond the SM~\cite{Kolb:1990}. Although
recent hints of possible WIMP signals may be encouraging, the lack of clear and convincing evidence for the discovery of WIMP dark matter
motivates consideration of dark matter candidates which deviate from the expectations of the WIMP paradigm.  A well-motivated
example is light dark matter (LDM), a class of dark matter candidates with masses typically $\sim 10 \mev - 10 \gev$.  LDM would
elastically scatter at direct detection experiments, with nuclear recoil energies which are relatively small and may be below the experimental
threshold, rendering them undetectable.  In this case, other experimental means aside from direct detection would be required to probe
DM-SM interactions.  For example, complimentary bounds on LDM scattering can be inferred from collider monojet
searches~\cite{Feng:2005gj,Goodman:2010,Birkedal:2004,Hooper:2010,Bai:2010,Rajaraman:2011,Fox:2012,Bai:2013,Agrawal:2013,
Goodman:2011-2,Bai:2011,Papucci:2014}.
Nonresonant LDM production at low energy $e^+/e^-$colliders can also be used to set model independent limits on electron scattering and, if
the LDM couples universally, nucleon scattering~\cite{Essig:2013}.

In this work, we consider the prospects for probing LDM-quark interactions through bounds on invisible decays of heavy quarkonium states at
colliders~\cite{Fayet:2007,Fayet:2010,Cotta:2013,Schmidt-Hoberg:2013hba,McElrath:2007}.  Such bounds have already been considered in a variety of contexts, including
$B$ and $D$ meson decays~\cite{Badin:2010}, and $\Upsilon$ decays into scalar LDM~\cite{Yeghiyan:2010,McKeen:2009rm}.  However, the constraints which one
can obtain on dark matter-quark
interactions depend in detail on the quantum numbers of the heavy meson, as well on the choice of final state
(i.e., $\rightarrow invisible$ or $\rightarrow \gamma +invisible$).
The angular momentum and $C/P$ transformation properties of the initial state (as well as the presence or absence of a photon in the
final state) together determine
which of the possible dark matter-quark interaction structures can participate in the decay process (and can thus be bounded by constraints
on invisible decays).
We consider invisible decays of the heavy quarkonium states $\Upsilon (1S)$ and $J / \Psi $, mesons with $J^{PC} = 1^{--}$.
As the quark constituents annihilate in an $s$-wave, the dependence of the meson decay matrix element on
the associated nonrelativistic bound state wavefunction is very simple and can be determined experimentally, with relatively little uncertainty.
Moreover, since the mesons which we consider each have a quark and anti-quark of the same flavor, the DM-SM interactions which
we introduce are not constrained by bounds on flavor-violation and, given universal quark coupling, can contribute to nucleon scattering.

At quark level, the matrix element relevant for meson decay ($\bar q q \rightarrow \bar X X$) is also relevant for
monojet/photon/$W$,$Z$ searches at the LHC~\cite{CMS:2012, CMS:2012tea,ATLAS:2012ky,ATLAS:2012,ATLAS:2013}.  As a result, these searches
will share many features, and event rates will have the same
dependence on the energy of the process.  Moreover, bounds arising from both invisible meson decay rates and LHC monojet
searches do not weaken as the dark matter mass decreases, in notable contrast to direct detection searches.
However, if the particle mediating LDM interactions is light, then the interaction is poorly approximated by a contact
operator for the purposes of LHC mono-anything searches,
and model independent bounds on the interaction strength can no longer be obtained. Even given a mediating particle massive
enough to warrant the use of the contact interaction approximation, the masses of LDM particles would be difficult to resolve at LHC searches due to
the large center-of-mass energy of the beam.
The bounds obtained from meson decays thus provide a unique handle on some dark matter interaction models, which is complementary
to the information provided by other search strategies.

In this paper, we use the limits on bound state decay widths to constrain the coupling of scalar, fermion or vector LDM to Standard Model quarks through
all contact operators of dimension six or lower.
In section II, we review the relevant effective contact interactions and calculate the resulting meson
decay rates and dark matter annihilation and scattering cross sections.
In section III we present constraints on all of the relevant dark matter-quark interaction structures
arising from $\Upsilon (1S)$ decay and Fermi gamma-ray searches
of dwarf spheroidal galaxies, and relate these constraints to those arising from direct detection experiments and LHC searches.
We conclude in section IV with a discussion of our results.

\section{Framework and Constraints}

We will consider a framework in which dark matter-quark interactions can be parametrized by a four-point contact effective
operator.  Such a structure can be written as an appropriate Lorentz contraction of a Standard Model quark bilinear and a
dark matter bilinear.  We are only interested in effective operators which can yield a non-zero matrix element when acting
on a $1^{--}$ meson state, such as the $\Upsilon (1S)$ or $J/\psi$; for such operators, the quark bilinear must be either
$ \bar q \gamma^i q$ or $ \bar q \sigma^{0 i} q$, where $i$ is a spatial index~\cite{Kumar:2013iva}.
The angular momentum quantum numbers of the quark/anti-quark bound states of interest are $S=1$, $L=0$, $J=1$.
In general, the final state need not have the same $C$ and $P$ transformation properties as the initial state (and thus need not have
the same $S$ and $L$ quantum numbers), but must have the
same total angular momentum $J$.  The effective operators\footnote{Henceforth, we refer to spin-0, spin-1/2 and spin-1 dark matter
fields with $\phi$, $X$ and $B^\mu$, respectively.  When describing a dark matter particle of arbitrary spin, we will use $X$.}
of dimension 6 or less which can have a non-zero matrix element with either an $\Upsilon (1S)$ or
$J/\psi$ initial state are listed in Table~\ref{tab:structs}~\cite{Kumar:2013iva} (the operators are labeled using the conventions
of~\cite{Kumar:2013iva} and~\cite{Goodman:2010}).
Note that for Majorana fermion dark matter, F6 is the only non-vanishing contact operator.

\begin{table}[hear]
\centering
\begin{tabular}{|c|c|c|c|c|}
\hline
 Name & Interaction Structure &  Annihilation  & Scattering \\
\hline
 F5  & $ (1/\Lambda^2) \bar X \gamma^\mu X \bar q \gamma_\mu q$ & Yes & SI \\
\hline
 F6  & $ (1/\Lambda^2) \bar X \gamma^\mu \gamma^5 X \bar q \gamma_\mu q$ & No & No \\
\hline
 F9  & $ (1/\Lambda^2) \bar X \sigma^{\mu \nu} X \bar q \sigma_{\mu \nu} q$ & Yes & SD \\
\hline
 F10  & $ (1/\Lambda^2) \bar X \sigma^{\mu \nu} \gamma^5 X \bar q \sigma_{\mu \nu} q$ & Yes & No \\
\hline
 S3  & $ (1/\Lambda^2) \imath Im (\phi^\dagger\partial _\mu \phi ) \bar q \gamma^\mu q$ & No & SI \\
\hline
 V3  & $ (1/\Lambda^2 ) \imath Im (B_\nu^\dagger\partial _\mu B^\nu ) \bar q \gamma^\mu q$ & No & SI \\
\hline
 V5  & $ (1/\Lambda ) (B_\mu^\dagger B_\nu - B_\nu^\dagger B_\mu ) \bar q \sigma^{\mu \nu} q$ & Yes & SD \\
\hline
 V6  & $ (1/\Lambda ) (B_\mu^\dagger B_\nu - B_\nu^\dagger B_\mu ) \bar q \sigma^{\mu \nu} \gamma^5 q$ & Yes & No \\
\hline
 V7  & $ (1/\Lambda^2) B_\nu^{(\dagger)} \partial^\nu B_\mu \bar q \gamma^\mu q$ & No & No \\
\hline
 V9  & $ (1/\Lambda^2) \epsilon^{ \mu \nu \rho \sigma} B_\nu^{(\dagger)} \partial_\rho B_\sigma \bar q \gamma_\mu q$ & No & No \\
\hline
\end{tabular}
\caption{Effective contact operators which can mediate the decay of a $J^{PC} = 1^{--}$
quarkonium bound state.
We also indicate if the operator can permit an $s$-wave dark matter initial state to annihilate to a quark/anti-quark
pair; if so, then a bound can also be set by indirect observations of photons originating from
dwarf spheroidal galaxies. Lastly, we indicate if the effective operator can mediate velocity-independent nucleon scattering
which is either spin-independent (SI) or spin-dependent (SD).}
\label{tab:structs}
\end{table}

The measured limits on the invisible
decay branching fractions could possibly be contaminated by decays to an invisible final state as well as a soft photon.  However,
contributions to the processes $\Upsilon (1S), J/\psi \rightarrow \bar X X \gamma$ would not arise from the operators we consider.  A contribution could
arise from the other dark matter-quark four-point operators which we do not consider, but the rate would be suppressed by an additional
factor of $\alpha$; to constrain such operators, it would be more fruitful to search directly for $\gamma +invisible $ decays.
The contribution of soft photon processes to fake decays was also considered in~\cite{Aditya:2012im}, where
the contribution was also small.

The meson decay rate matrix element can be determined by convolving the quark/antiquark annihilation ($\bar q q \rightarrow \bar X X$)
matrix element with the meson bound state wavefunction.
Using crossing symmetry, one can relate the quark annihilation matrix element to the matrix elements for
either dark matter annihilation ($\bar X X \rightarrow \bar q q$) or dark matter-nucleon scattering ($XN \rightarrow XN$).

The signals which can be observed at various experiments
depend on which quarks appear in the effective operator.  The $b$- and $c$-quark couplings will be relevant for the decay of
$\Upsilon$ or $J/\psi$ states, respectively,  whereas $u$-, $d$- and $s$-quark couplings are the most relevant for either dark matter
annihilation, dark matter-nucleon scattering, or dark matter production at the LHC.

\subsection{Bound State Decays}

The matrix element for the decay of a bound state is given by the convolution of the nonrelativistic bound state wavefunction with the
annihilation matrix element for a free quark/anti-quark pair.  Since we consider $s$-wave meson bound states in the nonrelativistic
approximation, this convolution depends only
on the value of the spatial wavefunction at the origin, $\psi (0)$.  The wavefunction at the origin can then be determined
from the well-measured decay branching fraction to $e^+ e^-$, yielding
\bea
{\cal B} ( \Upsilon (1S) \rightarrow e^{+} e^{-} ) = 16 \pi\alpha^2 Q_b^2 { | \psi_\Upsilon (0) |^2 \over \Gamma_\Upsilon M_\Upsilon^2 }
= 0.0238 \pm 0.0011 ,
\nonumber\\
{\cal B} ( J / \Psi  \rightarrow e^{+} e^{-} ) = 16 \pi \alpha^2 Q_c^2 { | \psi_{J/\psi} (0) |^2 \over \Gamma_{J/\psi} M_{J/\psi}^2 }
= 0.0594 \pm 0.0006 ,
\eea
with $M_\Upsilon = 9460.30 \pm 0.26 \mev$, $\Gamma_\Upsilon = 54.02 \pm 1.25 \kev$, $M_{J/\psi} = 3096.916 \pm 0.011 \mev$
and $\Gamma_{J/\psi} = 92.9 \pm 2.8 \kev$~\cite{PDG}. Note, we have ignored the contribution from $Z$,$h$-exchange; this
contribution is smaller than the uncertainties
in the measured branching fraction.
Searches for $\Upsilon$(1S) invisible decays have been performed by Belle~\cite{Belle:2007} and BaBar~\cite{BaBar:2009}
operating at the $\Upsilon (3S)$ resonance. They use the transition $\Upsilon(3S) \rightarrow\pi^{+}\pi^{-}\Upsilon (1S)$ to detect invisible
$\Upsilon (1S)$ decays and reconstruct the presence of the $\Upsilon$(1S) from the $\Upsilon(1S)$ peak in the recoil mass distribution, $M_{rec}$,
by tagging $\pi^{+}\pi^{-}$ pairs with kinematics
\bea
M_{rec}^{2} \equiv s + M_{\pi \pi }^{2}-2\sqrt{s}E^{*}_{\pi \pi },
\eea
where $M_{\pi \pi}$ is the invariant mass of the dipion system, $E^{*}_{\pi \pi}$ is the energy of the dipion system in the
center-of-mass (CM) frame of the $\Upsilon(3S)$, and $\sqrt{s} = 10.3552 \gev $ is $\Upsilon (3S)$ resonance energy.
Similar searches for invisible decays of $J/\Psi$ are based on the transition $\Psi(2S)\rightarrow\pi^{+}\pi^{-}J/\Psi$.
The 90\% CL constraints on branching fractions for invisible decays of $\Upsilon (1S)$ and $J / \Psi$, as
measured by BaBar and BES~\cite{BES:2008}, respectively, are
\bea
{\cal B} ( \Upsilon (1S) \rightarrow invisible ) < 3.0 \times 10^{-4},
\nonumber\\
{\cal B} ( J / \Psi  \rightarrow invisible ) < 7.2 \times 10^{-4} .
\eea
There is a Standard Model contribution to invisible bound state decays, namely, the decay of
a meson to $\bar \nu \nu$ via a $Z$ boson.
But these partial widths have been calculated and are negligible~\cite{Chang:1998}:
\bea
{ \cal B} ( \Upsilon (1S) \rightarrow \nu \bar \nu ) &=& 9.85 \times 10^{-6} ,
\nonumber\\
{ \cal B} ( J / \Psi \rightarrow \nu \bar \nu ) &=& 2.70 \times 10^{-8} .
\eea

For each contact operator, one
can calculate the branching fraction for the bound state to decay to dark matter in terms of
the bound state mass, the mediation scale, the dark matter mass, and
the branching fraction to $e^{+} / e^{-}$  (assuming $q=b$ for $\Upsilon (1S)$ decay, or $q=c$ for $J/\psi$ decay):
\bea
{\cal B}_{F 5}  (\bar X X )&=& { {\cal B} ( e^{+} e^{-} ) M^4 \over 16 \pi^2 \alpha^2 Q^2 \Lambda^4 }
\left( 1 - {4 m_X^2 \over M^2} \right)^{1/2} \left(1 + {2 m_X^2 \over M^2 } \right) ,
\nonumber\\
{\cal B}_{F 6}  (\bar X X )&=& { {\cal B} ( e^{+} e^{-} ) M^4 \over 16 \pi^2 \alpha^2 Q^2 \Lambda^4 } \left( 1 - {4 m_X^2 \over M^2} \right)^{3 / 2} ,
\nonumber\\
{\cal B}_{F 9}  (\bar X X )&=& { {\cal B} ( e^{+} e^{-} ) M^4 \over 8 \pi^2 \alpha^2 Q^2 \Lambda^4 }
\left( 1 - {4 m_X^2 \over M^2} \right)^{1/2} \left(1 + {8 m_X^2 \over M^2 } \right) ,
\nonumber\\
{\cal B}_{F 10}  (\bar X X )&=& { {\cal B} ( e^{+} e^{-}) M^4 \over 8 \pi^2 \alpha^2 Q^2 \Lambda^4 } \left( 1 - {4 m_X^2 \over M^2} \right)^{3 / 2} ,
\nonumber\\
{\cal B}_{S 3}  (\bar X X )&=& { {\cal B} ( e^{+} e^{-} ) M^4 \over 256 \pi^2 \alpha^2 Q^2 \Lambda^4 } \left( 1 - {4 m_X^2 \over M^2} \right)^{3 / 2} ,
\nonumber\\
{\cal B}_{V 3}  (\bar X X )&=& { {\cal B} ( e^{+} e^{-} ) M^4  \over 128 \pi^2 \alpha^2 Q^2 \Lambda^4 } \left( 1 - {4 m_X^2 \over M^2} \right)^{3/2}
\left(1 + { M^4 \over 8 m_X^4 } \left( 1 - {2 m_X^2 \over M^2} \right)^2 \right) ,
\nonumber\\
{\cal B}_{V 5}  (\bar X X )&=& { {\cal B} ( e^{+} e^{-} ) M^2 \over 16 \pi^2 \alpha^2 Q^2 \Lambda^2 } \left( 1 - {4 m_X^2 \over M^2} \right)^{3/2}
{M^2 \over m_X^2} \left( 1 + { M^2 \over 4 m_X^2} \right) ,
\nonumber\\
{\cal B}_{V 6}  (\bar X X )&=& { {\cal B} ( e^{+} e^{-} ) M^2 \over 16 \pi^2 \alpha^2 Q^2 \Lambda^2 }
\left( 1 - {4 m_X^2 \over M^2} \right)^{1/2} {M^2 \over m_X^2}  \left(1 + {2 m_X^2 \over M^2 } \right) ,
\nonumber\\
{\cal B}_{V 7}  (\bar X X )&=& {  {\cal B} ( e^{+} e^{-} ) M^4 \over 64 \pi^2 \alpha^2 Q^2 \Lambda^4 } \left( 1 - {4 m_X^2 \over M^2} \right)^{3/2}
{M^2 \over m_X^2} ,
\nonumber\\
{\cal B}_{V 9}  (\bar X X )&=& {{\cal B} ( e^{+} e^{-} ) M^4 \over 256 \pi^2 \alpha^2 Q^2 \Lambda^4 } \left( 1 - {4 m_X^2 \over M^2} \right)^{5/2}
{M^2 \over m_X^2} .
\eea
Note that for operator F6 we have written the branching fraction assuming that dark matter is a Dirac fermion.  If the dark matter were
instead a Majorana fermion, the branching fraction would be larger by a factor of $2$ (with a factor of $4$ arising from
the squared matrix element, and a factor of $1/2$ from the reduced phase space of the final state particles).
The result for operator S3, and the corresponding constraints, match that found in~\cite{Yeghiyan:2010}.

Note that if dark matter is spin-1,
the decay rates have terms which scale as $m_X^{-2}$ or $m_X^{-4}$.  These terms arise from final states if
either one or both of the dark matter particles is longitudinally polarized~\cite{Kumar:2013iva}.

In general, unitarity constrains the magnitude of the squared matrix element for both elastic and inelastic
scattering.  One may worry that, for small $m_X$, unitarity would require the presence of corrections which invalidate
the use of the contact operator approximation at tree-level; for example, if the dark matter is a
gauge boson which has become massive due to spontaneous breaking of a dark sector symmetry, then one may also need
to include additional diagrams involving the fields responsible for the spontaneous symmetry breaking.
However, the constraints from unitarity are
trivial when initial particles are at rest, because the elastic scattering cross-section is at threshold (for example,
see~\cite{Endo:2014mja}).  As a result, the contact operator approximation is consistent with unitarity for the
meson decay process in the non-relativistic bound state limit.  This approximation becomes more problematic as one
departs from the non-relativistic limit, but that analysis is beyond the scope of this work.

\subsection{Dark Matter Annihilation}

Applying crossing symmetry to the quark annihilation matrix elements yields the matrix element for dark matter annihilation
to quarks.  If the dark matter can annihilate from an $s$-wave initial state, then gamma ray observations can be used to set
limits on $\Lambda$.  But one should note that, although we have restricted ourselves to effective operators which have a
non-zero matrix element with an $s$-wave meson state, these operators need not necessarily have a non-zero matrix element
with an $s$-wave dark matter initial state.  Only five of these operators which we consider can permit unsuppressed dark matter
annihilation.  The corresponding annihilation cross-sections, at tree-level, are given by

\bea
\langle \sigma_A^{F 5} v \rangle &=& { 3 \over 2 \pi \Lambda^4  } \left( 1 - { m_q^2 \over m_X^2} \right)^{1/2}
\left( 2 m_X^2 + m_q^2 \right) ,
\nonumber\\
\langle \sigma_A^{F 9} v \rangle &=& { 6 \over \pi \Lambda^4  } \left( 1 - { m_q^2 \over m_X^2} \right)^{1/2}
\left( m_X^2 + 2 m_q^2 \right) ,
\nonumber\\
\langle \sigma_A^{F 10} v \rangle &=& { 6\over \pi  \Lambda^4  } \left( 1 - { m_q^2 \over m_X^2} \right)^{ 3 / 2}
m_X^2 ,
\nonumber\\
\langle \sigma_A^{V 5} v \rangle &=& { 2 \over 3 \pi  \Lambda^2  } \left( 1 - { m_q^2 \over m_X^2} \right)^{ 3 / 2}
 ,
\nonumber\\
\langle \sigma_A^{V 6} v \rangle &=& { 2 \over 3 \pi \Lambda^2  } \left( 1 - { m_q^2 \over m_X^2} \right)^{1/2}
\left( 1 +  {2 m_q^2 \over m_X^2} \right) ,
\eea

These cross sections can be bounded by a stacked analysis of the number of photons
arriving from dwarf spheroidal galaxies~\cite{Fermi:2011,Fermi:2013,Geringer:2011,Geringer:2012,Essig:2009}.
The number of photons
expected to result from dark matter annihilation is the product of an astrophysics-dependent
factor and a particle-physics dependent factor.  The astrophysics-dependent factor depends on the density profile
of the dwarf spheroidal galaxies, and can be estimated from the rotation curves of visible matter.
The particle-physics dependent factor can be expressed as
\bea
\Phi_{PP} &=& { \langle \sigma_A v \rangle \over 8 \pi m_X^2 } \int_{E_{thr}}^{m_X} \sum_f B_f {d N_f \over d E} d E ,
\label{eqn:gammaflux}
\eea
where $B_f$ is the branching ratio for dark matter to annihilate to a channel $f$ and $E_{thr} = 1 \gev$ is
the photon energy analysis threshold. $ d N_f / d E$ is the photon spectrum for a given dark matter
annihilation channel. Note that our bounds from dark matter annihilation assume universal quark coupling, with the strongest
contribution to the photon spectrum coming from the $u$- and $d$- quark channels.

\begin{figure}[hear]
\center\includegraphics[width=0.5 \textwidth]{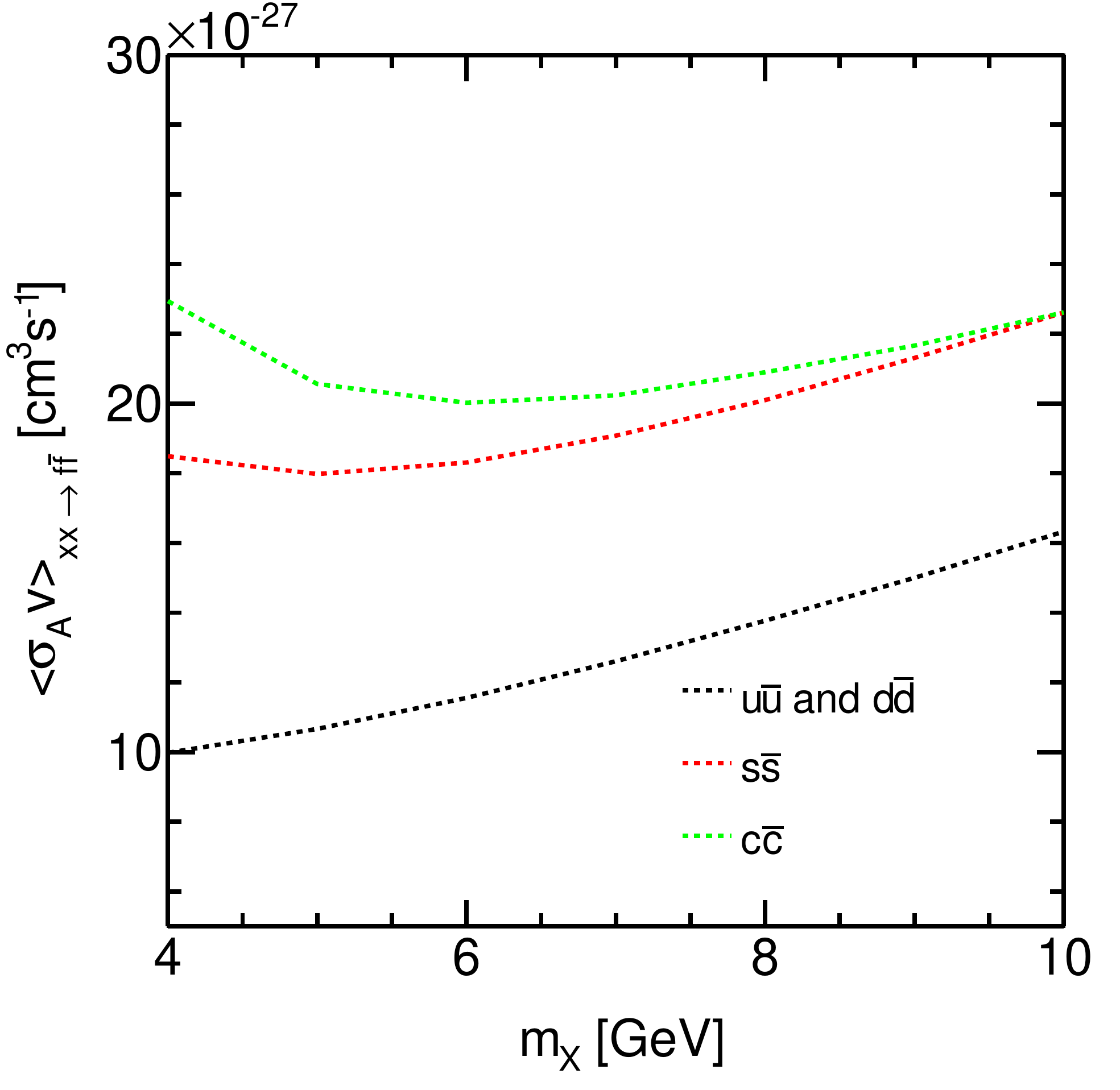}
\caption{Bounds on the annihilation cross section, $\langle \sigma_A v \rangle$, for dark matter of mass $m_X$
annihilating to quarks in dwarf spheroidal galaxies. Note that the results for annihilation to the $u \bar u $ and $d \bar d$
channels are visually identical.}
\label{fig:sigann}
\end{figure}

The $ 95 \% $ CL limit on $\Phi_{PP}$ arising from observations by Fermi-LAT of dwarf spheroidal galaxies is given by~\cite{Geringer:2011}
\bea
\Phi_{PP} < 5.0_{-4.5}^{+4.3} \times 10^{-30} \, {\rm cm}^3 \, {\rm s}^{-1} \gev^{-2},
\eea
where the asymmetric uncertainties are $95 \%$ CL systematic errors~\cite{Fermi:2011}, resulting from the uncertainty in
the mass density profiles of the various satellites.  This bound assumes that dark matter is its own anti-particle
(if the dark matter field is complex, as in the cases we consider,
this number increases by a factor of 2).
The associated bounds on the annihilation cross sections are plotted in Figure~\ref{fig:sigann}. For each
annihilation channel, the photon spectrum was produced by Pythia $6.403$~\cite{Pythia} and accounts for collinear photons
generated during the quark final state parton shower evolution.

The general shape of the bound on the annihilation cross-section is easily understood; for larger dark matter mass, the
bound strengthens with decreasing mass because of the resulting increase in the dark matter number density. At lower
masses the bound begins to sharply weaken because it is no longer possible to create
photons above the analysis threshold. We do not plot bounds on dark matter annihilation for $m_X \lsim 4 \gev$ since the energy of the final state quarks
becomes close to the hadronization scale.
Annihilation constraints for lighter dark matter would also be cut off by either, for the $\bar c c$ and $\bar s s$ channels,
the threshold to create quark/antiquark final states or, for the $\bar u u$ and $\bar d d$ channels, the analysis threshold.

Given any particle physics model for the dark matter annihilation cross section and final state branching
fractions, one can then determine if the model is consistent with data from Fermi-LAT~\cite{Kumar:2011}.
Note, however, that since all of the relevant operators are only non-vanishing when dark matter is
complex, it is consistent with this analysis for dark matter to be asymmetric.  In this case, only the
particle would be abundant at the current epoch, not the anti-particle; the bounds on dark
matter annihilation in dwarf spheroidal galaxies would thus necessarily be unconstraining.

It should be noted that, for $m_X \lsim {\cal O}(1)~\gev $,  $s$-wave dark matter annihilation  can be strongly constrained by anisotropies
in the cosmic microwave background (CMB) arising from the energy deposition caused by annihilation during recombination~\cite{Essig:2013goa}.
But since we do not assume any coupling between dark matter and leptons, the only relevant annihilation channels are $\bar X X \rightarrow \bar q q$.
For these final states, if $m_X \lsim 1~\gev$, it is more difficult to determine CMB constraints on LDM because of the effects of hadronization.
A full treatment of these constraints is thus beyond the scope of this work.  But one can note that, in general,
the constraints on the scale $\Lambda$ arising from CMB studies (as with gamma-ray searches of dwarf spheroidal
galaxies) either tend to become weaker as $m_X$ decreases, for fermionc dark matter, or strengthen less, relative to collider-based constraints, for vector dark matter.

\subsection{Nuclear Scattering}

Several of the effective operators which we consider will also yield velocity independent terms in
the associated dark matter-nucleon scattering matrix element.

Operators with vector quark bilinears will yield spin-independent scattering, with associated cross sections given by
\bea
\sigma_{SI}^{p,n} &=& { \mu_p^2 \over 32 \pi (2J_X+1) } \sum_{spins}
\left| \sum_q  { B_q^{p,n} \over m_X m_q} {\cal M}_{Xq \rightarrow Xq}  \right|^2   ,
\eea
where $\mu_p$ is the reduced mass of the dark matter-nucleon system and $J_X$ is
the dark matter spin.
The nucleon form factors associated with the vector quark bilinear are $B_u^p = B_d^n = 2$, $B_u^n = B_d^p = 1$, and
$B_{s,c,b,t}^{p,n}=0$~\cite{Ellis:2001}.

Similarly, operators with tensor quark bilinears will yield spin-dependent scattering, with associated cross sections given by
\bea
\sigma_{SD}^{p,n} &=& { \mu_p^2 \over 32 \pi (2J_X +1)} \sum_{spins}
\left| \sum_q  {\delta_q^{p,n} \over m_X m_q}
{\cal M}_{Xq \rightarrow Xq}  \right|^2 ,
\eea
The nucleon spin form factors $\delta_q^{p,n}$ can be extracted from data and are roughly given by
$\delta_u^p = 0.54_{-0.22}^{+0.09}$ and $\delta_d^p = - 0.23_{-0.16}^{+0.09}$~\cite{Fan:2010,Anselmino:2009}.  These form factors are
in slight disagreement with lattice calculations~\cite{Belanger:2008,Aoki:1997,LHPC:2002}, although using alternative form
factors would only change our scattering
cross sections by a factor of order unity.

Note that for contact operators which we consider, only the coupling to first-generation quarks is relevant for scattering.  In general,
dark matter can exhibit velocity-independent spin-independent scattering from nucleons through interactions with heavy quarks, but only if
the dark matter bilinear is a scalar.  We do not consider such operators, however, because they cannot contribute to the decay of a
$1^{--}$ meson.  Thus, the bounds on dark matter-quark interactions arising from $\Upsilon (1S)$ or $J/\psi$ decay can only be related
to bounds on dark matter-nucleon scattering if one makes a particular choice for the relative strength of dark matter coupling to light and
heavy quarks.

\section{Results}

We assume that dark matter couples to quarks only through a single effective operator, but with equal coupling
to all quark flavors.  An example of a model which would yield this effective operator realization would be the
case where the mediating particle was a massive vector boson for a new $U(1)$ symmetry under which all quarks have
equal charge and under which the dark matter is also charged (for example, this $U(1)$ could be a linear combination
of $U(1)_{baryon}$ and another $U(1)$ symmetry under which the dark matter is charged).  But, of course, other choices
are possible and can be well-motivated by other UV completions.  We focus on the case of equal couplings simply as a benchmark.
We note also that the coefficient of the effective contact operator can exhibit RG-running between the energy scale relevant
for meson decay and the scale relevant for nuclear scattering; however, this is expected to be a relatively small effect~\cite{Haisch:2013uaa,Crivellin:2014qxa}.

\subsection{Mediator Scale}

\begin{figure}[hear]
\includegraphics*[width=0.49 \textwidth]{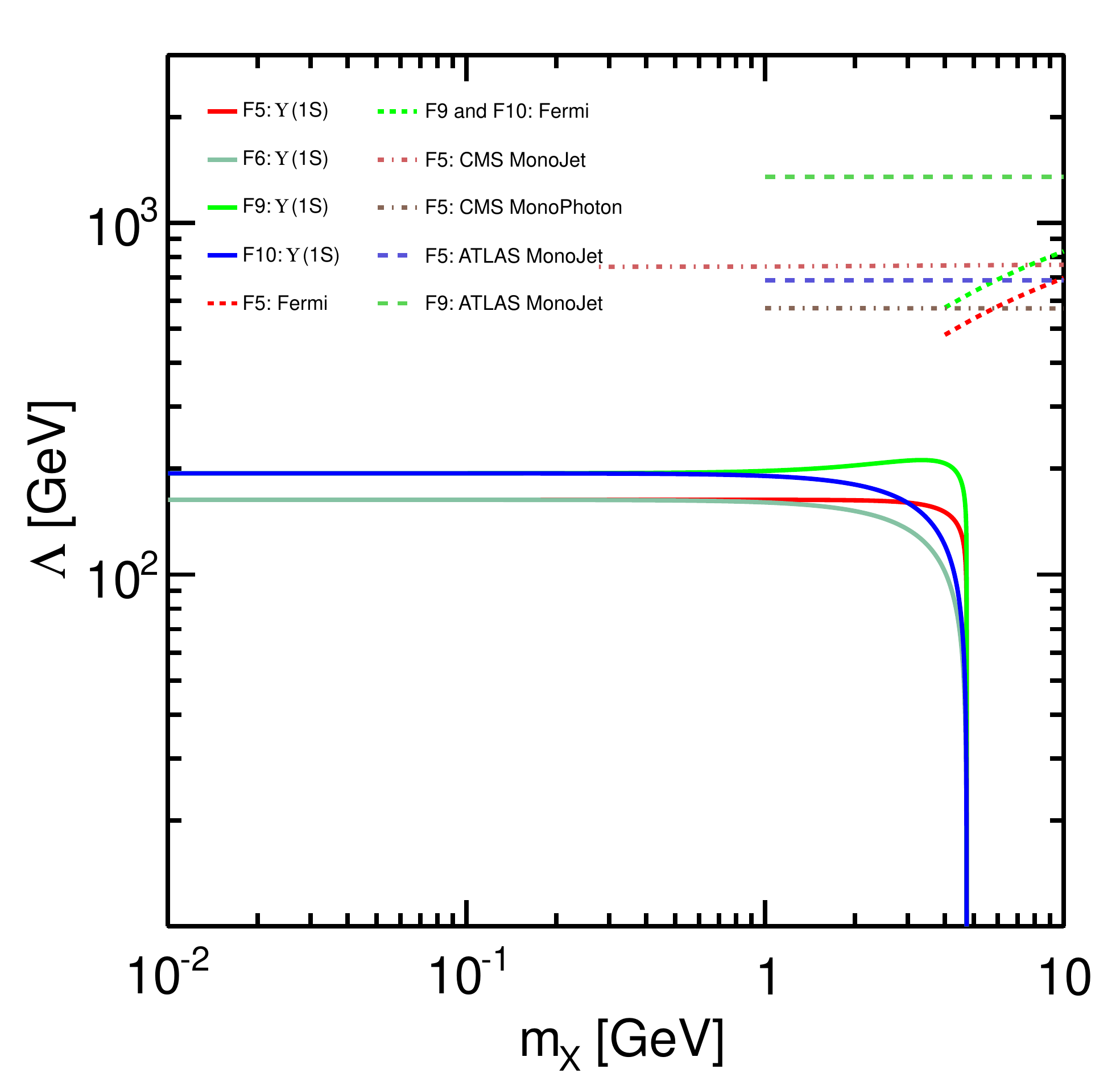}
\includegraphics*[width=0.49 \textwidth]{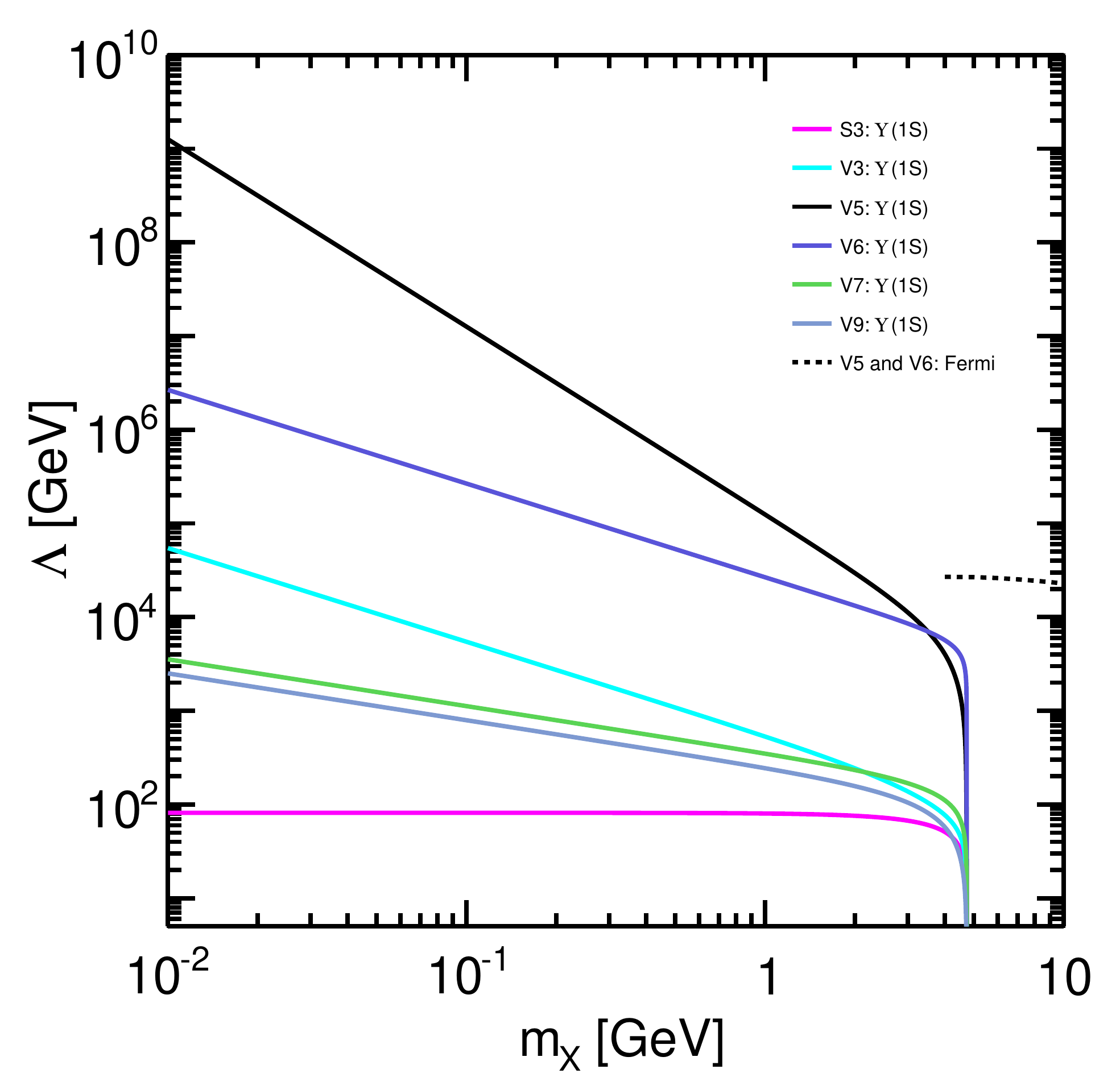}
\caption{Bounds on the mediator scale, $\Lambda$, for fermionic dark matter (left panel) and bosonic dark matter (right panel)
of mass $m_X$ arising from constraints on $\Upsilon (1S) \rightarrow nothing$ decays, from constraints on dark matter annihilation
to light quarks in dwarf spheroidal galaxies, and from monojet/photon searches at ATLAS~\cite{ATLAS:2012ky,ATLAS:2012} and
CMS~\cite{CMS:2012, CMS:2012tea}, as indicated.}
\label{fig:lambda}
\end{figure}

In Figure~\ref{fig:lambda} we plot bounds on $\Lambda$, as a function of $m_X$,
arising from limits on invisible $\Upsilon (1S)$ decays and from
dark matter annihilation in dwarf spheroidal galaxies.
Note that the bounds arising from $\Upsilon (1S)$ decays are sensitive only to dark matter coupling to $b$-quarks.  But for values
of $m_X$ for which $\Upsilon (1S)$ decays are kinematically allowed, dark matter annihilation to $\bar b b$ is kinematically
forbidden\footnote{For the case where $m_X > m_b$, bounds arising from radiative $\Upsilon$ decays at the LHC
have been discussed in~\cite{Cotta:2013,Schmidt-Hoberg:2013hba}.}; instead, the bounds on dark matter annihilation arise from dark matter couplings to the lighter quarks.
Note that
the mediator scale, $\Lambda$, is always larger than $\sim 10~\gev$, implying that the contact operator
approximation is valid at all of the colliders which are relevant for meson decay bounds.

Bounds on dark matter annihilation only arise for the operators which permit dark matter annihilation from an $s$-wave initial
state; dark matter annihilation through other operators is suppressed by factors of $v^2$, implying that current gamma ray observations
can provide no useful bounds.
In particular, if dark matter is spin-0, then dark matter annihilation is always $p$-wave if mediated by a contact operator
which can also mediate the decay of a $1^{--}$ meson, and it is thus unconstrained by gamma ray observations.
Bounds on dark matter annihilation in dwarf spheroidal galaxies are not constraining for $m_X \leq 1~\gev$,
because very light dark matter cannot produce any photons above the analysis threshold.  Bounds on the dark matter annihilation
cross section have systematic uncertainties related to the dark matter density profile; these uncertainties can weaken these bounds
by up to a factor of 2 or strengthen them by up to a factor of 10.  

We also plot bounds on the scale $\Lambda$ for each of the operators (for spin-1/2 dark matter) arising from searches
for monojet and monophoton production at the LHC~\cite{CMS:2012, CMS:2012tea,ATLAS:2012ky,ATLAS:2012} (similar bounds can also be
found for mono-$W$,$Z$ production~\cite{ATLAS:2013}).
These searches place bounds on the cross section for the process $pp \rightarrow XX+jet,\gamma$, where
dark matter interacts with Standard Model quarks through a contact operator.  As such, these bounds are very similar in spirit to
bounds on dark matter-quark couplings arising from meson decay, and in particular these bounds do not dramatically worsen for
very light dark matter.  However, it is important to note one significant difference.  These LHC monojet bounds are only valid if
the contact operator approximation is valid even at the energies of LHC processes, which requires the mediator mass to be ${\cal O}(\tev)$.
For lighter mediators, one cannot perform a model-independent operator analysis; although stringent bounds may be
possible~\cite{Busoni:2013lha,Papucci:2014iwa}, they depend on
the details of dark matter-quark interaction.  Of course, the bounds arising from meson decay are also only valid if the contact operator
approximation is valid, but in this case the relevant energy scale is the meson mass, which is ${\cal O}(1-10~\gev)$.  For mediators much
heavier than $\sim 10~\gev$, the bounds which arise from meson decay will be valid.

Dark matter with $m_X \gsim 10 \mev $ is heavy enough to be consistent with the cold dark matter paradigm.
Note, however, that our analysis does not require that dark matter is a thermal relic or that dark matter
annihilation be predominantly to Standard Model final states.
Instead, our focus is on constraints on dark matter interactions arising from astrophysical observations of the annihilation products
in the current epoch, not on constraints on the mechanism of dark matter generation.
Even if the the cross section for dark matter annihilation to visible matter is
small, there could be a large branching fraction for annihilation to the dark sector, evading indirect detection constraints but
allowing for a thermal relic density which can satisfy observational constraints.

\subsection{Complementary Dark Matter Scattering}

Of the contact operators which we consider, the ones which permit velocity-independent scattering are
F5 (SI), F9 (SD), S3 (SI), V3 (SI), and V5 (SD).  The corresponding total dark matter-proton scattering
cross sections are given by
\bea
\sigma_{SI}^{F5}
= { \mu_p^2 \over \pi \Lambda^4 } \left( B_u^p + B_d^p \right)^2 ,
\nonumber\\
\sigma_{SI}^{S3} = \sigma_{SI}^{V3}
= { \mu_p^2 \over 4 \pi \Lambda^4 } \left( B_u^p + B_d^p \right)^2 ,
\eea
\bea
\sigma_{SD}^{F9} &=& { 12 \mu_p^2 \over \pi \Lambda^4 }
\left( \delta_u^p + \delta_d^p \right)^2 ,
\nonumber\\
\sigma_{SD}^{V5} &=& { 2 \mu_p^2 \over  \pi \Lambda^2 m_X^2 }
\left( \delta_u^p + \delta_d^p \right)^2 .
\eea

\begin{figure}[hear]
\includegraphics*[width=0.49 \textwidth]{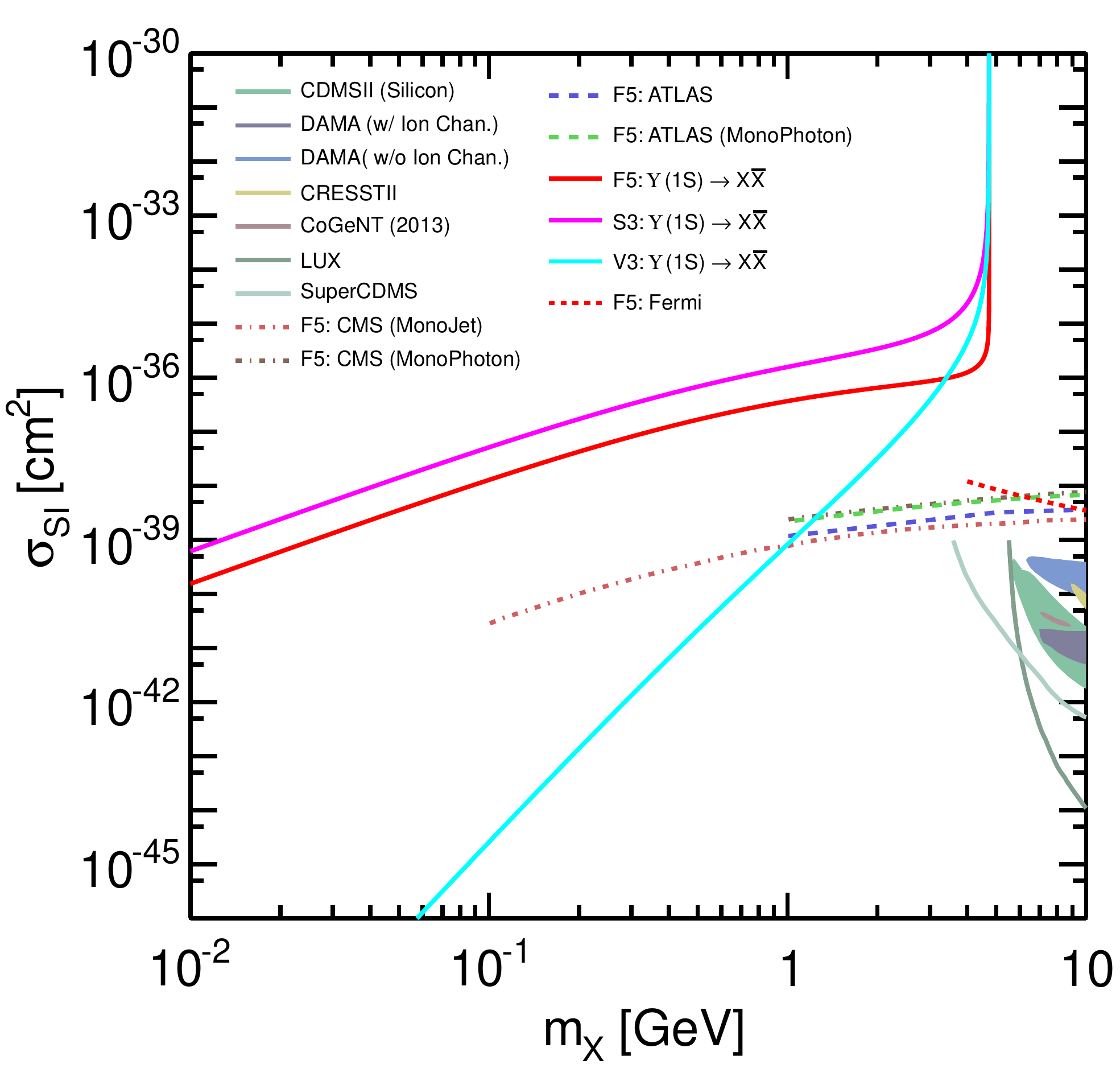}
\includegraphics*[width=0.49 \textwidth]{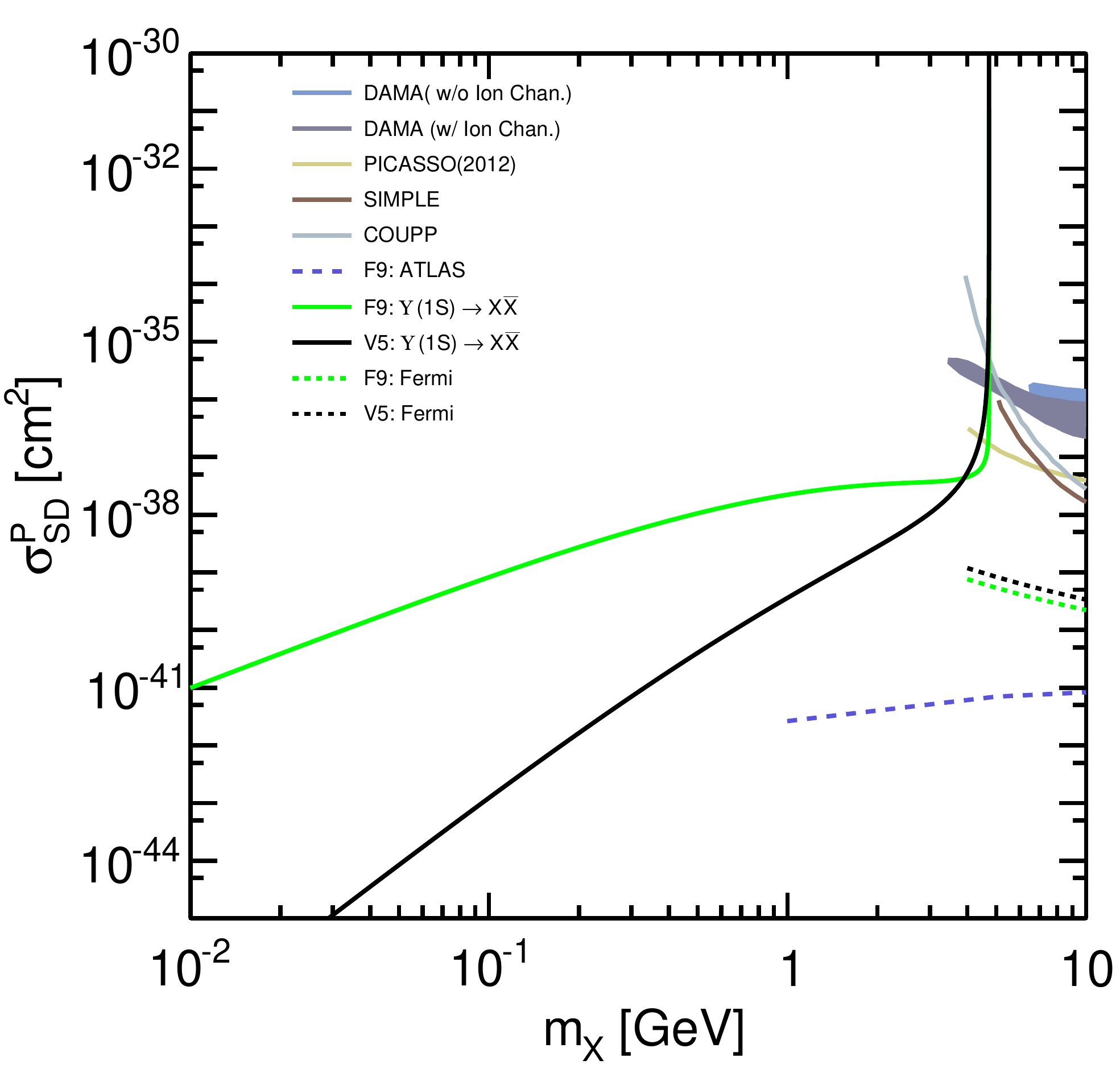}
\caption{
Bounds on the dark matter-proton spin-independent (left panel) and spin-dependent (right panel) scattering cross section
for dark matter of mass $m_X$ coupling universally to quarks through the indicated effective contact operator.
The labeled exclusion contours indicate 90\% CL bounds arising from limits on invisible decays of $\Upsilon (1S)$,
95\% CL bounds arising from Fermi-LAT constraints on dark matter annihilation in dwarf spheroidal galaxies,
and 90\% CL bounds arising from monojet searches (CMS~\cite{CMS:2012, CMS:2012tea} and ATLAS~\cite{ATLAS:2012ky}).
The DAMA/LIBRA~\cite{Savage:2008er}, CRESST II (95\% CL)~\cite{Angloher:2011uu}, CoGeNT~\cite{Aalseth:2012if} and
CDMS II(Silicon)~\cite{Agnese:2013rvf} 90\% CL signal regions
are also shown, as are the 90\% CL exclusion contours from
SuperCDMS~\cite{CDMS:2014}, LUX~\cite{Akerib:2013tjd}, SIMPLE~\cite{Felizardo:2011uw}, PICASSO~\cite{Archambault:2012pm},
and COUPP~\cite{Behnke:2012ys}.}
\label{fig:scattering}
\end{figure}

In Figure~\ref{fig:scattering}, we plot the bounds on spin-independent (left panel) and spin-dependent (right panel) scattering mediated by
each of the relevant operators.
We also plot 95\% CL bounds arising from Fermi-LAT constraints on dark matter annihilation in dwarf spheroidal galaxies
and 90\% CL bounds arising from monojet and monophoton searches (CMS~\cite{CMS:2012, CMS:2012tea} and ATLAS~\cite{ATLAS:2012ky,ATLAS:2012}).
The DAMA/LIBRA~\cite{Savage:2008er}, CRESST II (95\% CL)~\cite{Angloher:2011uu}, CoGeNT~\cite{Aalseth:2012if}, and CDMS II(Silicon)~\cite{Agnese:2013rvf}
90\% CL signal regions are also shown, as are the 90\% CL exclusion contours from SuperCDMS~\cite{CDMS:2014}, LUX~\cite{Akerib:2013tjd},
SIMPLE~\cite{Felizardo:2011uw}, PICASSO~\cite{Archambault:2012pm},
and COUPP~\cite{Behnke:2012ys}.

Various experiments are able to set bounds on DM-electron scattering~\cite{Essig:2013,Essig:2012}, but some
assumption of universal dark matter coupling to quarks and leptons is required to, in turn, bound dark matter-nucleon scattering.

Similar bounds on dark matter interactions can be obtained from bounds on $J / \Psi$ invisible decays, and are presented in the appendix.
While these bounds are weaker than those obtained from invisible $\Upsilon (1S)$ decays, they are valid for a larger range of mediator masses
($\gsim M_{J/\psi}$) and directly probe the couple of dark matter to $c$-quarks, thus providing non-trivial complementary constraints.
Note that the contact operator
approximation will begin to break down for mediator scales smaller than $\sim 10~\gev$.

\section{Conclusions}

We have presented bounds on dark matter-quark contact interactions which can be obtained from high luminosity $B$/charm-factories by
constraining decays
of the form $\Upsilon (1S), J/\psi \rightarrow \bar X X$.
These bounds on low mass dark matter probe a mass range significantly below the threshold of direct dark matter detection experiments
and complement bounds on dark matter interactions obtained from gamma ray searches of dwarf spheroidal galaxies and
from monojet/monophoton/mono-$W$,$Z$ searches at hadron colliders.

In particular, the effective interactions which permit decay of a $1^{--}$ meson state can also permit velocity-independent
dark matter-nucleon scattering (either spin-independent or spin-dependent).  For $m_X \sim 1-5~\gev$, the bounds obtained from meson decay can
thus potentially complement those obtained from direct detection experiments.  For the case of spin-independent scattering, direct detection
experiments already place bounds which well exceed those obtained from meson decay.  However, for spin-dependent scattering, bounds arising
from $\Upsilon (1S)$ decay via the F9 and V5 operators are comparable to those obtained from direct detection experiments.  This is not surprising,
as direct detection experiments typically have much weaker sensitivity to spin-dependent scattering, due to the lack of constructive interference
in coherent scattering.

Moreover, bounds on spin-1 dark matter interactions improve dramatically as $m_X$ decreases, because of the enhancement in the matrix element
which arises when the dark matter particles are longitudinally polarized.  Relating the suppression scale $\Lambda$ to the
mediator mass scale $m_{med.}$ and
coupling $g$ by $\Lambda \sim m_{med.} /g$, this implies that interactions between spin-1 dark matter and
quarks can be constrained even if the coupling is very weak.

We have seen that invisible quarkonium decays probe the same parton-level process ($\bar q q \rightarrow \bar X X$) as
monojet/photon/$W$,$Z$ searches at hadron colliders.  However, quarkonium decays provide complementary information,
allowing robust probes of models with relatively light mediators ($\gsim 10~\gev$) for which the contact operator
approximation would fail at the LHC.  It is also worth noting that, since the heavy quarkonium bound states are non-relativistic,
searches based on invisible heavy quarkonium decays readily distinguish between DM-SM interactions which vanish in the limit
of non-relativistic quarks and those which do not.  This probe thus nicely complements LHC searches, in which the partons
are highly relativistic.

A similar analysis can be performed of heavy quarkonium decays to a photon and missing energy;
although the set of relevant contact operators would be
different for such an analysis, those branching fractions are much more tightly constrained.
Improved bounds on the $\Upsilon (1S) \rightarrow nothing$ decay
rate from Belle II~\cite{Browder} and a factor
of $\sim 14$ enhancement in sensitivity to $J / \Psi \rightarrow nothing $ decay rate from BESIII~\cite{Harris} will
allow for even tighter constraints
on the interactions of low mass dark matter with Standard Model particles.

{\bf Acknowledgements}

We are grateful to Y.~G.~Aditya, T.~Browder, F.~Harris, D.~Marfatia, A.~Petrov, A.~Rajaraman, X.~Tata, S.~Vahsen, D.~Walker and A.~Wijangco for useful discussions.
This work is supported in part by Department of Energy grant  DE-SC0010504.

\appendix
\section{Constraints from $J / \Psi$ Decay}

\begin{figure}[hear]
\includegraphics*[width=0.49 \textwidth]{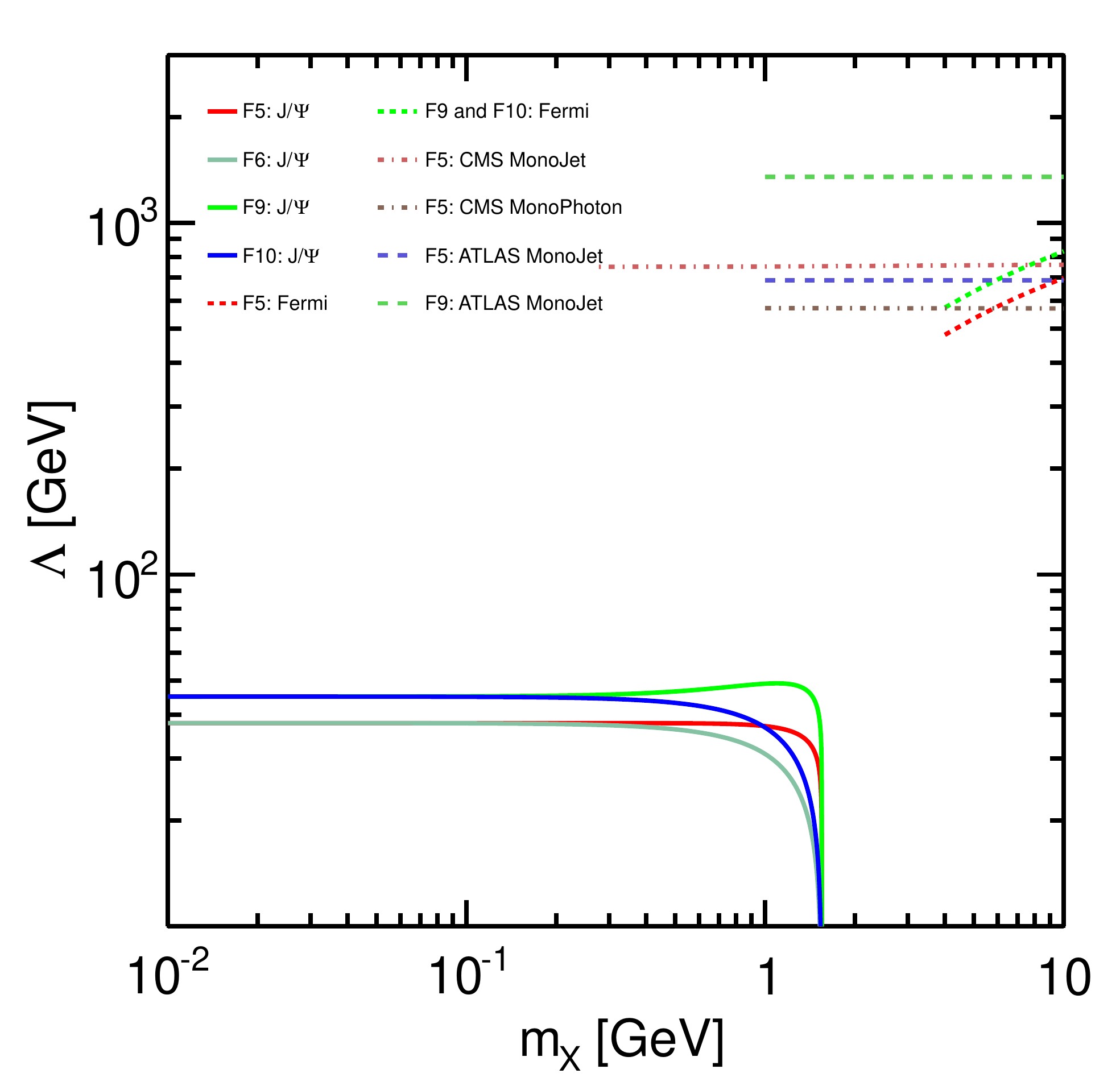}
\includegraphics*[width=0.49 \textwidth]{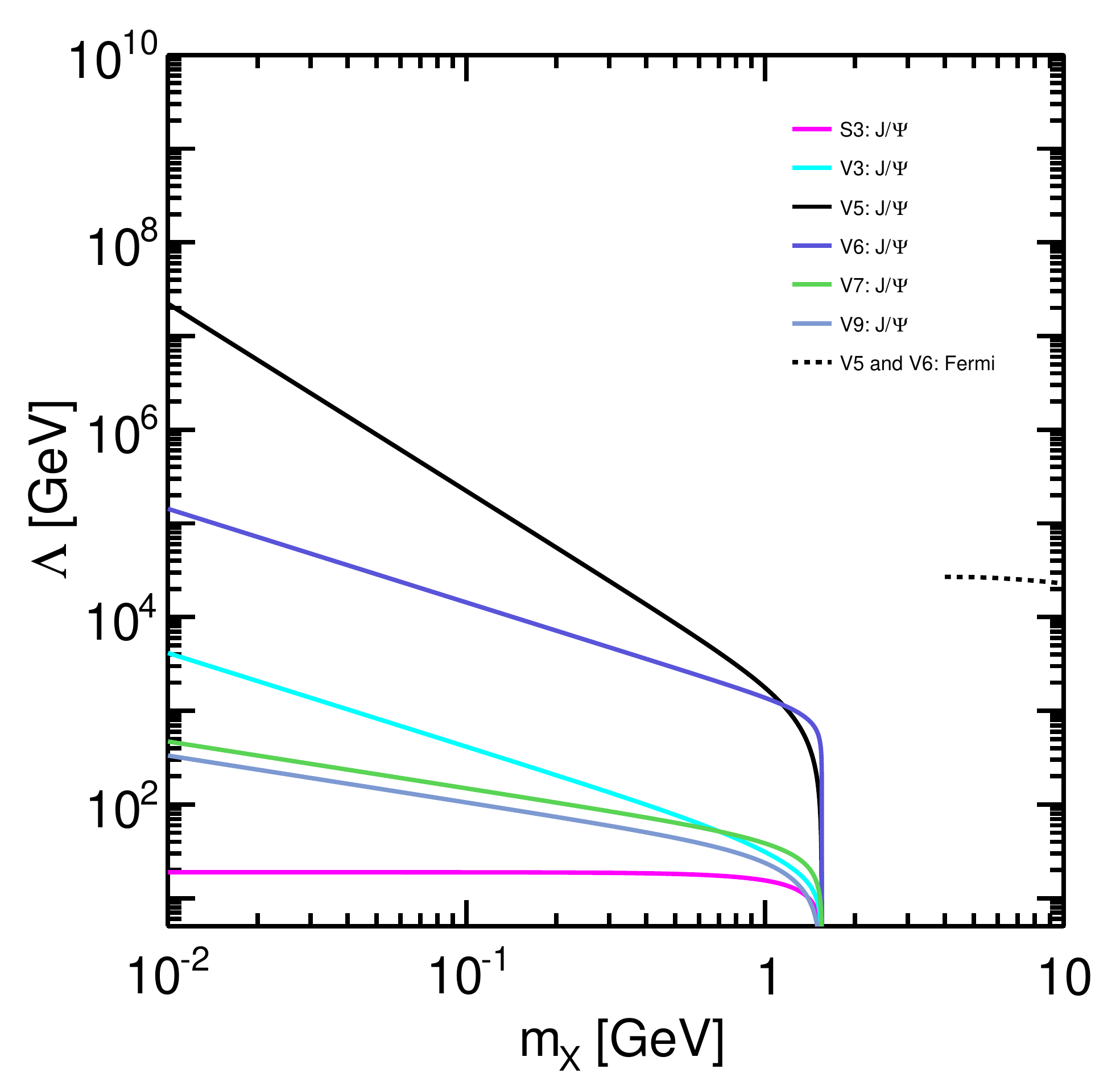}
\caption{Bounds on the mediator scale, $\Lambda$, for fermionic dark matter (left panel) and bosonic dark matter (right panel)
of mass $m_X$ arising from constraints on $J / \Psi \rightarrow nothing$ decays, from constraints on dark matter annihilation
to light quarks in dwarf spheroidal galaxies, and from monojet/photon searches at ATLAS~\cite{ATLAS:2012ky,ATLAS:2012} and
CMS~\cite{CMS:2012, CMS:2012tea}, as indicated.}
\label{fig:lambdaJpsi}
\end{figure}

In Figure~\ref{fig:lambdaJpsi}, we plot bounds on $\Lambda$, as a function of $m_X$
arising from limits on invisible $J / \Psi$ decays and from
dark matter annihilation in dwarf spheroidal galaxies.
We also plot bounds on the scale $\Lambda$ for each of the operators (for spin-1/2 dark matter) arising from searches
for monojet and monophoton production at the LHC~\cite{CMS:2012, CMS:2012tea,ATLAS:2012ky,ATLAS:2012} (similar bounds can also be
found for monophoton or mono-$W$,$Z$ production~\cite{ATLAS:2013}).
Note that the contact operator
approximation will begin to break down for mediator scales smaller than $\sim 10~\gev$.

\begin{figure}[hear]
\includegraphics*[width=0.49 \textwidth]{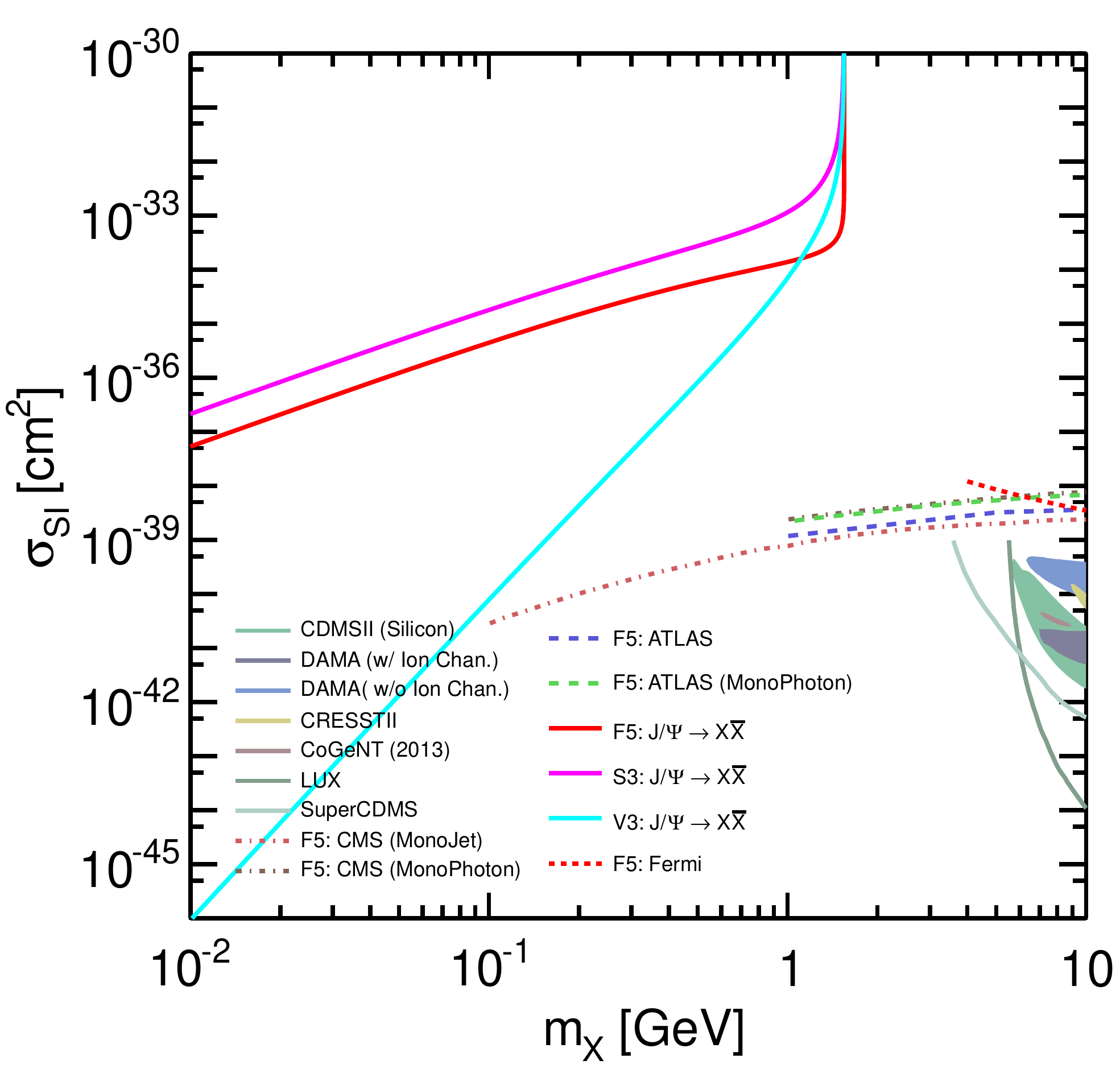}
\includegraphics*[width=0.49 \textwidth]{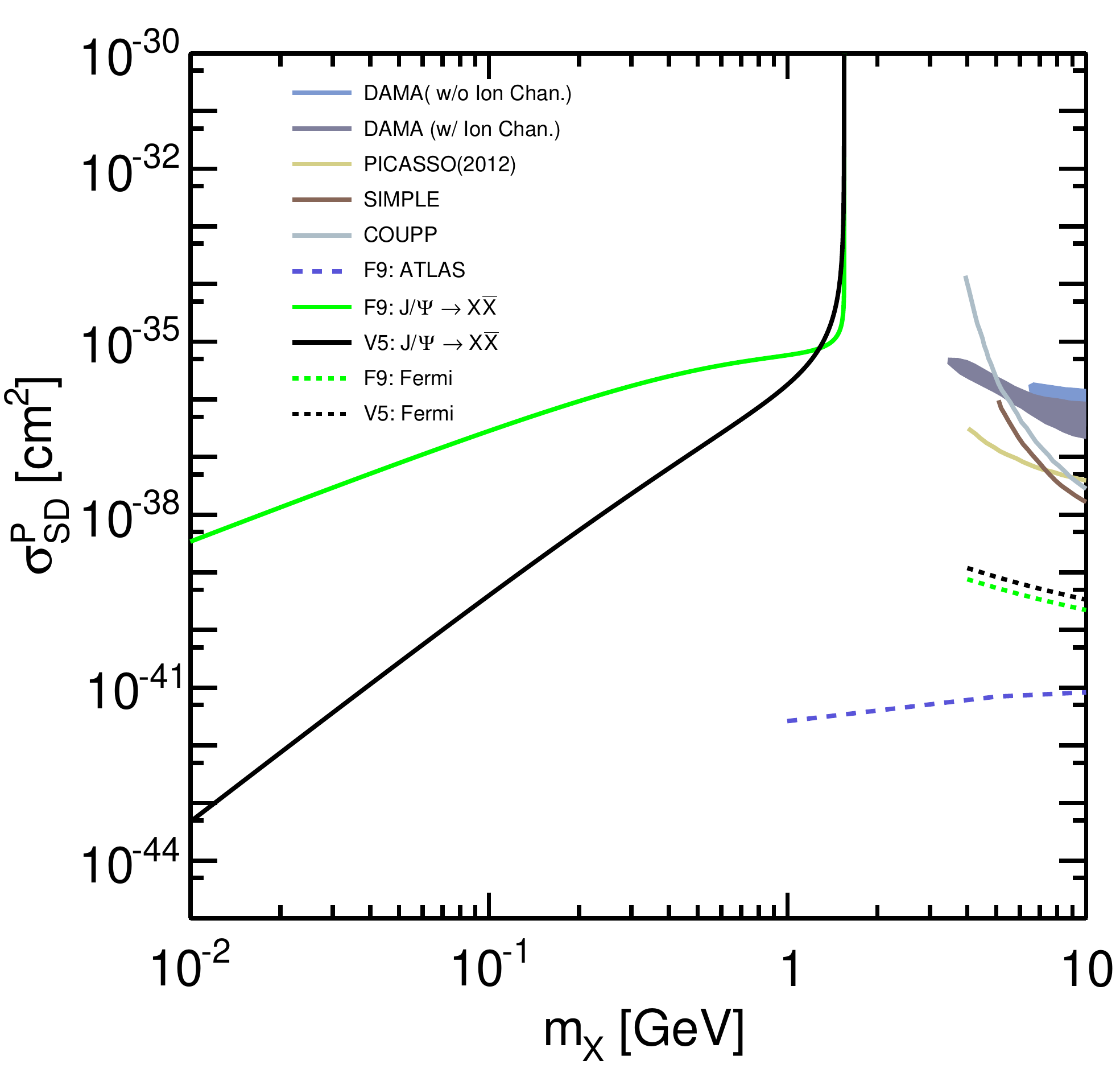}
\caption{
Bounds on the dark matter-proton spin-independent (left panel) and spin-dependent (right panel) scattering cross section
for dark matter of mass $m_X$ coupling universally to quarks through the indicated effective contact operator.
The labeled exclusion contours indicate 90\% CL bounds arising from limits on invisible decays of $J / \Psi$,
95 \% CL bounds arising from Fermi-LAT constraints on dark matter annihilation in dwarf spheroidal galaxies,
and 90 \% CL bounds arising from monojet searches (CMS~\cite{CMS:2012, CMS:2012tea} and ATLAS~\cite{ATLAS:2012ky}).
The DAMA/LIBRA~\cite{Savage:2008er}, CRESST II (95 \% CL)~\cite{Angloher:2011uu}, CoGeNT~\cite{Aalseth:2012if} and CDMS II(Silicon)~\cite{Agnese:2013rvf} 90 \% CL
signal regions
are also shown, as are the 90 \% CL exclusion countours from
SuperCDMS~\cite{CDMS:2014}, LUX~\cite{Akerib:2013tjd}, SIMPLE~\cite{Felizardo:2011uw}, PICASSO~\cite{Archambault:2012pm},
and COUPP~\cite{Behnke:2012ys}.}
\label{fig:scatteringJpsi}
\end{figure}

In Figure~\ref{fig:scatteringJpsi}, we plot the bounds on spin-independent (left panel) and spin-dependent (right panel) scattering mediated by
each of the relevant operators.
We also plot 95\% CL bounds arising from Fermi-LAT constraints on dark matter annihilation in dwarf spheroidal galaxies
and 90\% CL bounds arising from monojet and monophoton searches (CMS~\cite{CMS:2012, CMS:2012tea} and ATLAS~\cite{ATLAS:2012ky,ATLAS:2012}).
The DAMA/LIBRA~\cite{Savage:2008er}, CRESST II (95\% CL)~\cite{Angloher:2011uu}, CoGeNT~\cite{Aalseth:2012if}, and CDMS II(Silicon)~\cite{Agnese:2013rvf} 90\% CL
signal regions are also shown, as are the 90\% CL exclusion contours from SuperCDMS~\cite{CDMS:2014}, LUX~\cite{Akerib:2013tjd},
SIMPLE~\cite{Felizardo:2011uw}, PICASSO~\cite{Archambault:2012pm}, and COUPP~\cite{Behnke:2012ys}.

\end{document}